\documentclass[conference]{IEEEtran}
\IEEEoverridecommandlockouts
\usepackage{amsmath,amsfonts}
\usepackage{algorithmic}
\usepackage{algorithm}
\usepackage{array}
\usepackage[caption=false,font=normalsize,labelfont=sf,textfont=sf]{subfig}
\usepackage{textcomp}
\usepackage{stfloats}
\usepackage{url}
\usepackage{fontawesome}
\usepackage{verbatim}
\usepackage{graphicx}
\usepackage{cite}
\usepackage{tcolorbox}
\usepackage{multirow}
\usepackage{makecell}
\usepackage{booktabs}
\usepackage{hyperref}
\usepackage[utf8]{inputenc}
\usepackage{xcolor}
\usepackage[top=0.76in, bottom=1.04in, left=0.675in, right=0.675in]{geometry}
\begin{document}

\title{End-Cloud Collaboration Framework for Advanced AI Customer Service in E-commerce}

\author{
    \IEEEauthorblockN{
        Liangyu Teng\textsuperscript{1,4},
        Yang Liu\textsuperscript{2$^\ast$},
        Jing Liu\textsuperscript{3},
        Liang Song\textsuperscript{1,4$^\ast$\thanks{$^\ast$Corresponding authors.}}
    }

    \IEEEauthorblockA{
        \textsuperscript{1}Academy for Engineering \& Technology, Fudan University, Shanghai, China  \\
        \textsuperscript{2}Department of Computer Science, University of Toronto, Toronto, Canada\\
        \textsuperscript{3}Department of Electrical and Computer Engineering, The University of British Columbia, Vancouver, Canada\\
        \textsuperscript{4}Innovation Platform for Academicians of Hainan Province, Haikou, Hainan, China\\
        \texttt{lyteng24@m.fudan.edu.cn}, \texttt{yangliu@cs.toronto.edu}, \texttt{jingliu@ece.ubc.ca}, \texttt{songl@fudan.edu.cn}
    }
}

\markboth{Journal of \LaTeX\ Class Files,~Vol.~14, No.~8, July~2024}%
{Shell \MakeLowercase{\textit{et al.}}: End-Cloud Collaboration Framework for Advanced AI Customer Service in E-commerce}

\maketitle

\begin{abstract}
In recent years, the e-commerce industry has seen a rapid increase in the demand for advanced AI-driven customer service solutions. Traditional cloud-based models face limitations in terms of latency, personalized services, and privacy concerns. Furthermore, end devices often lack the computational resources to deploy large AI models effectively.
In this paper, we propose an innovative End-Cloud Collaboration (ECC) framework for advanced AI customer service in e-commerce. This framework integrates the advantages of large cloud models and mid/small-sized end models by deeply exploring the generalization potential of cloud models and effectively utilizing the computing power resources of terminal chips, alleviating the strain on computing resources to some extent. Specifically, the large cloud model acts as a teacher, guiding and promoting the learning of the end model, which significantly reduces the end model's reliance on large-scale, high-quality data and thereby addresses the data bottleneck in traditional end model training, offering a new paradigm for the rapid deployment of industry applications.
Additionally, we introduce an online evolutive learning strategy that enables the end model to continuously iterate and upgrade based on guidance from the cloud model and real-time user feedback. This strategy ensures that the model can flexibly adapt to the rapid changes in application scenarios while avoiding the uploading of sensitive information by performing local fine-tuning, achieving the dual goals of privacy protection and personalized service. %
To conclude, we implement in-depth corpus collection (e.g., data organization, cleaning, and preprocessing) and train an ECC-based industry-specific model for e-commerce customer service. Our ECC framework not only improves the response efficiency and service quality but also establishes a practical benchmark for customer service model development.
\end{abstract}

\begin{IEEEkeywords}
Large language model, end-cloud collaboration, e-commence customer service, online evolutive learning.
\end{IEEEkeywords}

\section{Introduction}
\label{sec:introduction}
Deep learning is an effective techique widely used in the field of Artificial Intelligence (AI) \cite{lecun2015deep}. By introducing multiple layers of neural networks, deep learning has been able to capture high-level abstractions in data and has been applied to various field including computer vision \cite{liu2023amp}, natural language processing and so on \cite{liu2023distributional}. In recent years, with the advent of Large Language Models (LLMs) such as GPT-3 \cite{brown2020language}, Gemini \cite{team2023gemini}, %
ChatGLM \cite{du2022glm} and GPT-4 \cite{achiam2023gpt}, scaling laws \cite{kaplan2020scaling} have become increasingly evident, demonstrating that as model parameters scale up, both the performance and generalization abilities of the models significantly improve. The trend of increasingly powerful comprehensive capabilities of models has catalyzed extensive utilization of these models in research across diverse industries including manufacturing \cite{zhou2024causalkgpt}, education \cite{kasneci2023chatgpt}, healthcare \cite{thirunavukarasu2023large}, %
and finance \cite{li2023large}.

The e-commerce domain is an important application scenario for LLMs \cite{li2024ecomgpt}. Well-trained e-commerce LLMs can act as intelligent customer service systems, offering personalized services that enhance customer satisfaction and drive sales. These systems provide round-the-clock support, rapid responses, and cost-efficiency. As a result, the use of LLMs in e-commerce customer service has become a research hotspot.

Leveraging LLMs for customer service in e-commerce has emerged as a novel application due to their powerful generalization capabilities and extensive knowledge. However, this approach faces challenges. According to \cite{wei2022emergent}, LLMs' human-like text generation arises from emergent abilities requiring large model sizes, leading to significant computational costs. Mainstream models, with billions to trillions of parameters, demand extensive computational resources. For instance, training GPT-4 costs over 10 million USD and requires more than 10,000 high-performance GPUs, making it impractical to train and deploy on end devices. Additionally, general LLMs lack specialized domain data, hindering their direct application in certain domains. Increasing parameter counts also heighten security and privacy risks due to immature understanding of LLM mechanisms.

To address the aforementioned problems, it is necessary to fundamentally shift the technical approach of LLMs by integrating them with terminals. The End-Cloud Collaboration (ECC) technology \cite{liu2024networking} offers a new perspective for research in the field of LLMs, becoming a new paradigm for large model applications on the end. In this paradigm, end models can accurately understand user needs and provide timely, personalized responses. Meanwhile, the large models in the cloud are adept at handling complex problems, taking over when end models are unable to perform complex tasks, thus meeting users' deeper needs.%

This paper constructs an ECC framework for Chinese e-commerce customer service, leveraging large cloud models to train mid/small-sized end models. We use the Gemini 1.5 pro model in the cloud to generate supervised information for training the end model, ChatGLM3-6B, which is suitable for consumer-grade GPUs. The end model ChatGLM3-6B is trained on Chinese customer service dialogues and cloud-provided supervised information using fine-tuning strategies like Prefix-Tuning \cite{li2021prefix}, P-Tuning-v2 \cite{liu2022p}, and LoRA \cite{hulora}. Additionally, the model can be evaluated either by human beings or by evualtion models and then self-optimize based on feedback from humans and the cloud model, achieving online evolutive learning.

The contributions of this paper are as follows.
\begin{itemize}
  \item We propose an ECC framework for e-commerce customer service, integrating cloud and end models to enhance deployment efficiency and address data bottlenecks.
  \item We introduce a dynamic online evolutive learning, which allow end models to adapt in real-time based on cloud guidance and user feedback, ensuring privacy and personalization.
  \item We develop a customized model fine-tuning method, improving service quality with a high-quality, targeted corpus dataset.
\end{itemize}

\section{Related Work}
\label{sec:relatedwork}

\subsection{End-Cloud Collaboration}

Edge computing addresses cloud computing's limitations in latency, bandwidth, and privacy by moving computation closer to the data source, thus reducing latency and bandwidth needs. However, it also faces constraints like limited computational resources and storage. To address these limitations, researchers have proposed the concept of edge-cloud collaboration \cite{yao2022edge,liu2024generalized}, which combines the strengths of both to create a more efficient and scalable platform. This paper adapts this concept to focus on end devices and cloud computing, termed end-cloud collaboration, leveraging the computational resources of both to enable real-time, efficient, and secure AI applications.

A classical example of edge-cloud collaboration is Federated Learning (FL) \cite{mcmahan2017communication}.
Through FL, end devices can learn from each other and improve the global model without compromising data privacy.
End-cloud collaboration has been shown to be effective in improving the performance of machine learning models and reducing the danger of data leakage.

\subsection{Large Language Models}
Large Language Models (LLMs) have been shown to achieve state-of-the-art performance on a wide range of natural language processing tasks, such as 
text generation and machine translation. LLMs are typically trained on large amounts of text data using unsupervised learning techniques, such as autoregressive language modeling and masked language modeling.

Nowadays, the most popular LLMs are based on Transformer \cite{vaswani2017attention}, which uses self-attention mechanisms to capture long-range dependencies in text data. Transformer has been shown to be highly effective in capturing complex patterns in text data and has been used in many pre-trained language models, such as BERT \cite{devlin2019bert}, GPT-3 \cite{brown2020language}, etc. %
Especially, ChatGPT demonstrated to people from all walks of life the powerful text generation capabilities of LLMs, which is the basis of many AI applications including chatbots.

\subsection{Fine-tuning Technicals}
In the field of natural language processing, the current mainstream paradigm is to pre-train LLMs on large-scale text corpora and then fine-tune them on specific tasks using supervised learning techniques. Fine-tuning involves training a pre-trained language model on a specific dataset to improve its performance on a specific task. Compared with the pre-trained model without fine-tuning, the fine-tuned model can achieve better performance on the specific task \cite{ding2023parameter}.

The traditional approach of fine-tuning, also known as full fine-tuning, is to train the entire model on the specific dataset. This approach requires training and storing a separate model for each task, leading to high computational costs, large storage requirements, and a significant need for labeled data. To address these challenges, researchers have proposed Parameter-Efficient Fine-tuning Techniques (PEFT), which aim to improve the performance of LLMs on specific tasks using a smaller dataset and fewer computational resources \cite{liu2023pre}. Methods such as Prefix-Tuning \cite{li2021prefix}, P-Tuning-v2 \cite{liu2022p}, and LoRA \cite{hulora} are examples of PEFT techniques.

\section{End-Cloud Collaboration Framework for Advanced E-commerce Customer Service Model}
\label{sec:framework}
In this paper, we design and implement an innovative ECC framework that leverages the power of cloud LLMs (in our experiment, we choose Gemini 1.5 pro) and the flexibility of the lightweight mid/small-sized model (e.g., ChatGLM3-6B) to jointly build an efficient and responsive AI IoT solution. %
This framework enhances interaction between the end and cloud, improving service intelligence and data processing efficiency.

The large model in the cloud serves as the knowledge center and data processing engine, extracting abstract knowledge from vast amounts of data to generate high-quality datasets for specific tasks. The mid/small-sized end model, deployed locally, is trained by both a natural dataset and the generated dataset. It provides real-time interaction and service execution, optimizing based on user feedback and cloud guidance.

\begin{figure}[t]
  \centering
  \includegraphics[width=\columnwidth]{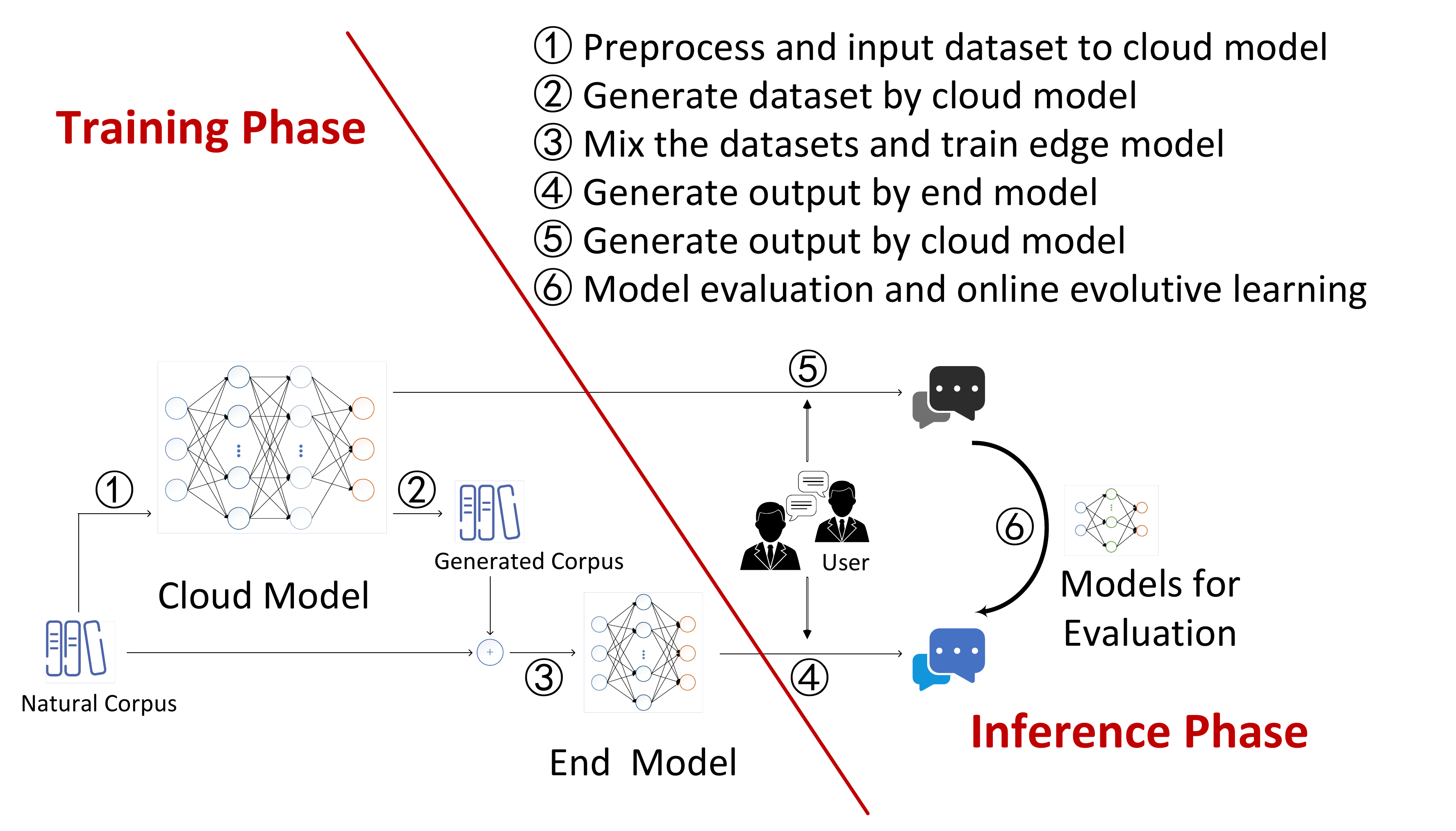}
  \caption{ECC framework for advanced e-commerce customer service model.}
  \label{fig:framework}
  \vspace{-5mm}
\end{figure}

As shown in Fig.~\ref{fig:framework}, the ECC framework consists of two main components: the cloud model, the end model. The cloud model is responsible for generating high-quality datasets and supervisory information for the end model. The end model is responsible for real-time interaction with users and providing personalized services. During the training phase, the cloud model generates high-quality datasets, which are used to train the end model. During the inference phase, the end model interacts with users in real-time and provides personalized services based on the training results.

To evaluate the model and make the model adjust to the users' demand continuously, the framework introduces a real-time feedback mechanism. We adopt a dual-track strategy of manual review and automated evaluation. The manual review by human-beings assesses whether the model's responses match user expectations, while automated evaluation leverages the cloud model which is adapted for e-commerce customer service via Prompt Engineering (PE). Specific details of PE are discussed in Section~\ref{sec:experiment}. Here, automated evaluation scores the end model by comparing its output to the output of the cloud model after PE, as well as assessing the degree of match between user inputs and model outputs. To be specific, we use a pre-trained model SimBERT \cite{simbert} which is based on BERT \cite{devlin2019bert} and can be used to calculate the similarity between the output text of the end model and of the cloud model. We used another pre-trained model RocketQA \cite{qu2020rocketqa} to calculate the degree of matching between the user inputs and the model outputs. After getting these two scores, we add these scores in a specific way to get the final score as follows:
\begin{equation}
  \small
  \text{Final Score} = \alpha \times \text{SimBERT Score} + 
  (1 - \alpha) \times \text{RocketQA Score},
\end{equation}
where $\alpha$ is hyperparameter to balance the two scores and $ \alpha \in [0, 1] $. The final score is then used to guide the end model to adjust its output results.

Notably, when the score of the cloud model under the RocketQA model is detected to be lower than a preset threshold $\Theta$, it implies that there is a poor match between the model's response and the user's original query. In this case, we believe that it is more appropriate to directly compare the semantic similarity (SimBERT score) of the model outputs, since the low match is due to the fact that the RocketQA model itself has not been trained on e-commerce domain data.

Based on the evaluation results, the framework adjusts the poor-quality output of the end model so that the end model can continuously self-optimize based on user interactions and the guidance of the cloud model, achieving online evolution and improving the level of personalized service. Specifically, when the end model's output dissatisfies the user's demand (e.g., gets a low score), the same query by the user will be input into the cloud model. The output of the cloud model will then become a pseudo label and will be used to train the end model.

Through this framework, the customer service model can further enhance user experience and personalized service. The end model possesses quick response and online evolutive capabilities, allowing customers to enjoy almost instantaneous and highly personalized interactions. The model can dynamically adjust based on historical interactions and user feedback, providing service suggestions and solutions that better meet user needs, thereby significantly improving user satisfaction. Moreover, the design of the ECC framework ensures that sensitive data processing can be performed locally on end devices. This effectively mitigates security risks and privacy breaches during data transmission, promoting data security and privacy protection.

\section{Experiment}
\label{sec:experiment}
\subsection{Datasets}
We mainly used two datasets, Taobao and Jing Dong e-commerce conversation datasets respectively. Taobao and Jing Dong are the two largest e-commerce platforms in China, and their conversation datasets are widely used in the field of Chinese e-commerce customer service.

Taobao E-commerce Dialogue Corpus (ECD) \cite{zhang2018modeling} consists of 1 million session-response pairs for training, 10,000 pairs for validation, and 10,000 pairs for testing.
Similar to ECD, Jing Dong Dialogue Corpus (JDDC) \cite{chen2020jddc} is also a widely used Chinese e-commerce conversation dataset with more than 1 million multi-turn dialogues, 20 million utterances, and 150 million words, which contains conversations about after-sales topics between users and customer service staffs in E-commerce scenario.

Table~\ref{tab:taobaoexample} shows an example of a dialogue in the ECD after preprocessing and translating to English. It demonstrates the characteristics of briefness and colloquial language typical in e-commerce domain dialogues.

\begin{table}[t]
  \caption{Dialogue example of ECD}\label{tab:taobaoexample}
  \vspace{-2mm}
  \begin{tcolorbox}[colback=blue!5!white,colframe=black,width=0.48\textwidth,title={A multi-turn dialogue example of ECD}]
  \scriptsize
  \faUser: Wait, what snack do I get after subscribing?\\
  \faCommentsO: The snack is given randomly, sir, like beef tendon or beef jerky. You'd better post a screenshot showing that you've subscribed our shop.I'll get you a discount coupon for your next dried mango purchase.\vspace{2mm}\\
  \faUser: Like this?\\
  \faCommentsO: You need to subscribe our shop, not the product, sir.\vspace{2mm}\\
  \faUser: Where do I subscribe the shop? I'm confused.\\
  \faCommentsO: Uh, go to our shop, and there is a follow button on the upper right corner, sir.\vspace{2mm}\\
  \faUser: Is this right?\\
  \faCommentsO: Yes, that's right. You can get a coupon for 3 yuan off when you spend 23 yuan on dried mango, dear sir. You can use it next time.\vspace{2mm}\\
  \faUser: Oh, okay.\\
  \faCommentsO: The coupon is valid until August 31. Thanks for supporting our shop.
  \end{tcolorbox}
  \vspace{-5mm}
\end{table}

Besides, we also use the cloud model, Gemini 1.5 pro, to generate a high-quality dataset for training the end model. We constructed the dataset by feeding the user input portion of the above dataset into the cloud LLM and using LLM's responses as label. Since Gemini 1.5 pro is a generalized model and does not directly generate statements that match the e-commerce style, we take advantage of the few-shot \cite{brown2020language} and in-context learning ability \cite{dong2022survey} of LLM to enable it to generate replies that are more in line with e-commerce scenarios by using the prompt shown in Table~\ref{tab:prompt} and the examples of the generative dataset are shown in Table~\ref{tab:response}. Here, N in Table~\ref{tab:prompt} represents the number of question-answer pairs in the prompt, and we set it 500 in our experiment. \faUser \  and \faCommentsO \  denoted the customer and the proposed e-commerce service model, respectively.

\begin{table}[t]
  \caption{Prompt on cloud model}\label{tab:prompt}
  \vspace{-2mm}
  \begin{tcolorbox}[colback=blue!5!white,colframe=black,width=0.48\textwidth,title={Prompt (N denotes the number of the dialogue)}]
  \scriptsize
  Assuming that you are an e-commerce customer service agent who is able to answer the specialized knowledge in e-commerce, has good service consciousness, and is able to deal with the customer's request properly, you are asked to play the role of an e-commerce customer service agent in order to deal with the customers in the following.
  \vspace{2mm}

  Here are some examples for your reference.
  \vspace{2mm}

  \begin{itemize}
    \item \faUser: Question No.1.
    \item \textbf{Customer Service}: Answer No.1.
    \item \faUser: Question No.2.
    \item \textbf{Customer Service}: Answer No.2.\\
    \vspace{2mm}
    \textbf\dots
    \item \faUser: Question No.N.
    \item \textbf{Customer Service}: Answer No.N.
  \end{itemize}
  \end{tcolorbox}
  \vspace{-5mm}
\end{table}

\begin{table}[t]
  \caption{Response of Gemini 1.5 pro}\label{tab:response}
  \vspace{-2mm}
  \begin{tcolorbox}[colback=blue!5!white,colframe=black,width=0.48\textwidth,title={Response of Gemini 1.5 pro before and after prompt engineering}]
  \scriptsize
  \faUser: Hello\\
  \textcolor{blue}{\textbf{Origin Answer by Gemini}: Hello! Nice to meet you! What would you like to talk about?}\\
  \textcolor{red}{\textbf{Current Answer by Gemini}: Hello}\vspace{2mm}\\
  \faUser: Can I choose the delivery service?\\
  \textcolor{blue}{\textbf{Origin Answer by Gemini}: I'm sorry, I need more information to understand what you mean. Can you tell me what you want to deliver, where you want it delivered, or which courier company you want to ask? Please provide more details so I can help you better.}\\
  \textcolor{red}{\textbf{Current Answer by Gemini}: Sir, we use ZTO Express by default.}\vspace{2mm}\\
  \faUser: 520401029636\\
  \textcolor{blue}{\textbf{Origin Answer by Gemini}: The number you provided, 520401029636, doesn't seem to have any specific meaning or context on its own.} (This sentence is originally written in English, all the rest is in Chinese, and has been translated into English.)\\
  \textcolor{red}{\textbf{Current Answer by Gemini}: This one is 20x20 cm, made of pure cotton, and can be used both wet and dry.}\\
  \end{tcolorbox}
  \vspace{-5mm}
\end{table}

\subsection{Environment and Parameter Setup}
The experiment is conducted on a server with 1 NVIDIA RTX3090 GPU with 24GB of memory. The server runs on Ubuntu 20.04, with Python 3.11, PyTorch 2.2.1 installed. Table~\ref{tab:parameters} shows the parameters for training and evaluating the end model ChatGLM3-6B.

\begin{table}[t]
  \centering
  \caption{Parameters of the end model ChatGLM3-6B under different fine-tuning methods}\label{tab:parameters}
  \resizebox{0.98\linewidth}{!}{
    \begin{tabular}{cccc}
    \toprule
    \textbf{Parameters} & \textbf{Prefix-Tuning} & \textbf{P-Tuning v2}  & \textbf{LoRA} \\
    \midrule
    \multicolumn{4}{c}{\textit{General Training Parameters}} \\
    \midrule
    Max Input Length & 256 & 256 & 256 \\
    Max Output Length & 512 & 512 & 512 \\
    Fine-Tuning Steps & 30000 & 30000 & 30000 \\
    Learning Rate & 5e-5 & 5e-5 & 5e-5 \\
    Per Device Train Batch Size & 1 & 1 & 1 \\
    \midrule
    \multicolumn{4}{c}{\textit{Specific Training Parameters}} \\
    \midrule
    num\_virtual\_tokens & 128 & 128 & / \\
    Rank r & / & / & 8 \\
    Alpha & / & / & 32 \\
    LoRA Dropout & / & / & 0.1 \\
    \midrule
    \multicolumn{4}{c}{\textit{Evaluation Parameters}} \\
    \midrule
    max\_length & 8192 & 8192 & 8192 \\
    $top_p$ & 0.8 & 0.8 & 0.8 \\
    temperature & 0.6 & 0.6 & 0.6 \\
    \bottomrule
    \end{tabular}
  }
  \vspace{-2mm}
\end{table}

\subsection{Experimental Results}
\begin{figure}[t]
  \centering
  \includegraphics[width=0.45\textwidth]{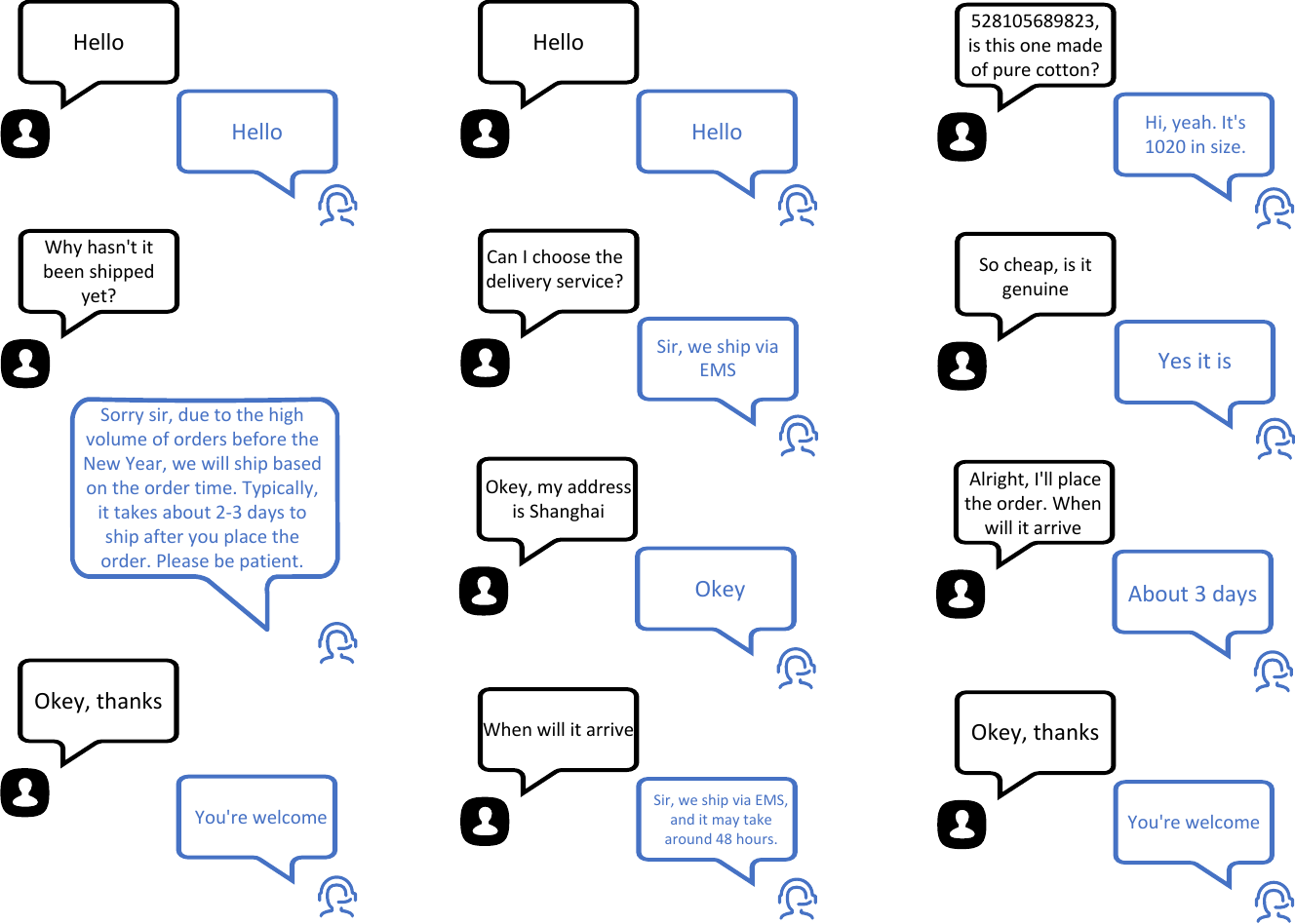}
  \caption{Examples of dialogue results from the end model ChatGLM3-6B.}
  \label{fig:conversation}
  \vspace{-5mm}
\end{figure}

After 15,000 steps of fine-tuning using LoRA, we get a fine-tuned model and test it through the conversations shown in Fig.~\ref{fig:conversation}. As shown in the figure, the model has been able to understand the characteristics of e-commerce customer service, and can respond to customer questions with contextual information.

Besides, we've compared the end model ChatGLM3-6B gotten from different fine-tuning methods and different steps in Table~\ref{tab:responseofglm}. Obviously, the model fine-tuned with LoRA works better in generating the e-commerce style text.

\begin{table}[!h]
  \caption{Response of ChatGLM3-6B}\label{tab:responseofglm}
  \vspace{-2mm}
  \begin{tcolorbox}[colback=blue!5!white,colframe=black,width=0.48\textwidth,title={Response of ChatGLM3-6B gotten from different methods and steps}]
  \scriptsize
  \faUser: Hello\\
  \textcolor{blue}{\textbf{ChatGLM3-6B}: Hello! I'm the AI assistant ChatGLM3-6B, nice to meet you, feel free to ask me any questions.}\\
  \textcolor{orange}{\textbf{Prefix-Tuning (step=3000)}: Hello! I'm the AI assistant ChatGLM3-6B, nice to meet you, feel free to ask me any questions.}\\
  \textcolor{green!50!black}{\textbf{P-Tuning v2 (step=3000)}: Hello! I'm the AI assistant ChatGLM3-6B, nice to meet you, feel free to ask me any questions.}\\
  \textcolor{pink}{\textbf{LoRA (step=3000)}: Hello}\\
  \textcolor{purple}{\textbf{LoRA (step=5000)}: Hello}\vspace{3mm}\\
  \faUser: Can I choose the delivery service?\\
  \textcolor{blue}{\textbf{ChatGLM3-6B}: Sure. As an artificial intelligence assistant, I can recommend some delivery companies for you to choose. In China, the common delivery companies are SF, Yuantong, Shentong, Yunda, Zhongtong, Bestway and so on. You can choose the right delivery company according to your needs and budget. If you need to know more about delivery companies, please feel free to let me know.}\\
  \textcolor{orange}{\textbf{Prefix-Tuning (step=3000)}: Are you inquiring about the option of a delivery company? Yes, you can choose a delivery company to send your items. Different delivery companies may have different range of services, prices, speed and reliability, so you can choose according to your needs. Usually, you can choose from the official website of the delivery company, or you can inquire about it by phone or online customer service.}\\
  \textcolor{green!50!black}{\textbf{P-Tuning v2 (step=3000)}: Of course you can, as an artificial intelligence assistant, I will recommend some delivery companies for you. In our country, SF, Yuantong, Shentong, Yunda, Zhongtong, Bestway and other delivery companies are very popular choices. You can choose the right delivery company according to your needs and budget. If you have other questions or need help, please feel free to let me know.}\\
  \textcolor{pink}{\textbf{LoRA (step=3000)}: Of course, dear customer}\\
  \textcolor{purple}{\textbf{LoRA (step=5000)}: Dear customer, you can choose the delivery service.}\vspace{3mm}\\
  \faUser: 520401029636\\
  \textcolor{blue}{\textbf{ChatGLM3-6B}: Hello! This looks like a string of numbers, what can I do for you please?}\\
  \textcolor{orange}{\textbf{Prefix-Tuning (step=3000)}: Hello, according to the number sequence you provided
  “520401029636”, I am unable to provide the appropriate answer or information at this time. Please provide a more detailed question or background information so that I can better assist you. Thanks!}\\
  \textcolor{green!50!black}{\textbf{P-Tuning v2 (step=3000)}: I'm sorry, I cannot understand the number sequence you provided “520401029636”. Please provide more information or background so that I can better answer your question.}\\
  \textcolor{pink}{\textbf{LoRA (step=3000)}: Hello}\\
  \textcolor{purple}{\textbf{LoRA (step=5000)}: Hello}\\
  \end{tcolorbox}
  \vspace{-4mm}
\end{table}

\begin{figure}[t]
  \centering
  \includegraphics[width=0.49\textwidth]{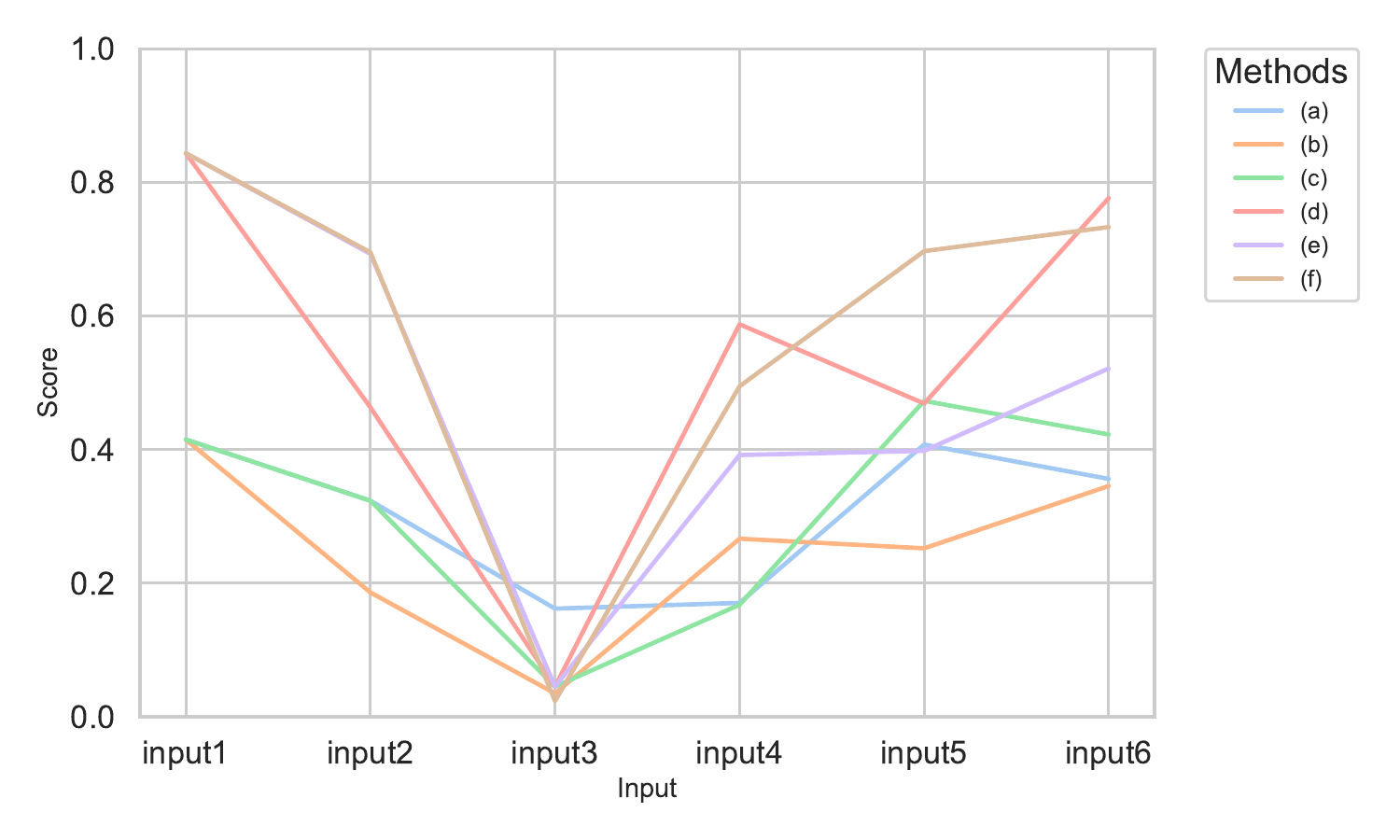}
  \caption{Visualization results of the end model ChatGLM3-6B under different fine-tuning methods. Here, (a) represents the original model, (b) represents the model fine-tuned with Prefix-Tuning after 3,000 steps, (c) represents the model fine-tuned with P-Tuning v2 after 3,000 steps, (d) represents the model fine-tuned with LoRA after 3,000 steps, (e) represents the model fine-tuned with LoRA after 5,000 steps, and (f) represents the model fine-tuned with LoRA after 15,000 steps.  Besides, the input 1-5 are the same as the ones in Table~\ref{tab:results} with the same order.}
  \label{fig:visualresult}
  \vspace{-4mm}
\end{figure}

To quantify the performance of the end model ChatGLM3-6B under different fine-tuning methods, we execute the automated evaluation in our EEC framework by setting $\alpha = 0.8$ and $\Theta = 0.2$. The results are shown in Table~\ref{tab:results} and Fig.~\ref{fig:visualresult}.

\begin{table}[t]
  \centering
  \caption{Evaluation results of the end model ChatGLM3-6B under different fine-tuning methods}\label{tab:results}
  \resizebox{0.49\textwidth}{!}{
    \begin{tabular}{c@{\hspace{0.4cm}}|@{\hspace{0.2cm}}c@{\hspace{0.2cm}}c@{\hspace{0.2cm}}c@{\hspace{0.2cm}}c@{\hspace{0.2cm}}c@{\hspace{0.2cm}}c}
    \toprule
    \multirow{2}{*}{User Input} & \multirow{2}{*}{Original Model} & Prefix-Tuning  & P-Tuning v2 & LoRA & LoRA & LoRA\\
    \small
    & & 3000 steps & 3000 steps & 3000 steps & 5000 steps &  15000 steps \\
    \midrule
    Hello & 0.415 & 0.415 & 0.415 &\bf 0.844 &\bf 0.844 &\bf 0.844 \\
    \midrule
    Can I choose the delivery service? & 0.324 & 0.186 & 0.324 & 0.463 & 0.693 &\bf 0.695 \\
    \midrule
    520401029636 &\bf 0.162 & 0.035 & 0.045 & 0.045 & 0.045 & 0.024 \\
    \midrule
    Can I get a discount? & 0.171 & 0.267 & 0.168 &\bf 0.588 & 0.392 & 0.495 \\
    \midrule
    \makecell{This is so frustrating. I've been \\ waiting for this delivery since yesterday \\ and haven't dared to leave the house.} & 0.408 & 0.253 & 0.473 & 0.469 & 0.398 &\bf 0.697 \\
    \midrule
    Will you ship the order if I purchase it today? & 0.356 & 0.346 & 0.423 &\bf 0.777 & 0.522 & 0.733 \\
    \bottomrule
    \end{tabular}
  }
  \vspace{-4mm}
\end{table}

It can be seen that under the same training iteration period, the LoRA method shows better performance than Prefix-Tuning and P-Tuning v2, which is not only more effective, but also faster in convergence. In addition, the result of the automated evaluation is consistent with the manual review (our subjective evaluation based on the samples in Table \ref{tab:responseofglm}), demonstrating the reasonableness and validity of our evaluation method.

\section{Conclusion}
\label{sec:conclusion}
In this paper, we propose and implement an ECC framework and apply it in the field of e-commerce customer service. This framework successfully integrates the large language model Gemini 1.5 pro with the mid-sized model ChatGLM3-6B, addressing the issues of traditional customer service's lack of personalized service and the difficulty of quickly adapting to market changes. The ECC framework leverages the strengths of both cloud-based and end-side models to create a highly flexible and efficient system.

The cloud model serves as the knowledge center and data processing engine, generating high-quality datasets optimized for specific tasks to ensure the accuracy and practicality of the outputs of the end model. It also serves as an evaluation criterion for the end-side model by comparing the similarity between the output of the end model and the output text of the cloud model. The end-side model, on the other hand, acts as a fast-responding intelligent agent and is deployed on local devices to perform real-time computations. It is trained using both natural datasets and the datasets generated by the cloud model. This model focuses on real-time interaction and service execution, adapting and optimizing its performance based on user feedback and guidance from the cloud model, i.e., a method of constantly fine-tuning itself using the output data from the cloud model as labels when users are dissatisfied with the response from the end-side model, thus enhancing the level of personalized service. Through this mechanism, we achieves online evolutive learning of the end model and promotes the intelligent upgrading of e-commerce customer service systems.

The ECC framework not only improves response speed and personalization in customer service but also ensures that sensitive data is processed locally on end devices. This design significantly reduces security risks and privacy issues during data transmission, enhancing data security and privacy protection.

In conclusion, the ECC framework not only enhances the efficiency and responsiveness of e-commerce customer service but also sets the stage for future advancements in AI-driven customer service systems. Future work will focus on further optimizing the ECC framework, exploring its application in other domains such as healthcare, finance, and education, and enhancing its adaptability to a wider range of business scenarios. We anticipate that the ECC framework will play a pivotal role in the evolution of intelligent customer service systems, fostering innovation and improving user satisfaction across various industries, including potential advancements in multimodal capabilities and multi-agent systems, thereby further enriching the customer service landscape.

\section*{Acknowledgments}
This work is supported in part by Fudan Institute on Networking Systems of AI and the Specific Research Fund of the Innovation Platform for Academicians of Hainan Province under Grant YSPTZX202314.
\bibliography{refer}

\vfill

\end{document}